\documentclass[preprint,aps]{revtex4}
\usepackage{epsfig,amsmath,graphics}
\newcommand{\be}{\begin{equation}}
\newcommand{\ee}{\end{equation}}
\newcommand{\bea}{\begin{eqnarray}}
\newcommand{\eea}{\end{eqnarray}}
\begin{document}


\title{Schr\"odinger equation for current carrying states}


\author{D.S. Kosov}

\email{kosov@theochem.uni-frankfurt.de}

\affiliation{Institute of Physical and Theoretical Chemistry, \\ 
J. W. Goethe University, Marie-Curie-Str. 11, D-60439 Frankfurt, Germany}

\begin{abstract}
Schr\"odinger equation with given,  {\it a priori } 
known current  is formulated. A non-zero current density
is maintained in the quantum system via a subsidiary condition 
imposed by vector, local Lagrange multiplier. Constrained 
minimization of the total energy on the manifold of an arbitrary  
current density topology results into a non-linear self-consistent 
Schr\"odinger equation. The applications to electronic transport 
in two-terminal molecular devices are developed and new macroscopic 
definition of a molecular current-voltage characteristic 
is proposed. The Landauer formula for the conductance of
an ideal one-dimensional lead is obtained within the approach. 
The method is examined by modeling of current carrying 
states of one-dimensional harmonic  oscillator.
\end{abstract}


\maketitle

\newpage

\section{Introduction}

The  remarkable miniaturization of conventional 
microelectronic devices  has been witnessed 
over the last decades. If the miniaturization trend
is to continue, elements of microelectronic circuits, 
e.g. transistors and contacts, will soon shrink
to single molecules or even atoms.
In the past few years, there have been considerable advances towards 
fabrications and experimental studies of mono-molecular electronic 
devices \cite{joachim2000}.
This activity has been largely spurred on by development of experimental 
techniques to form atomic scale electrical contacts such as scanning tunneling 
microscopy \cite{mirkin92} and mechanically controllable break 
junctions \cite{ruiten96}. Measurements
of current-voltage (I-V) characteristics have been recently performed on
a single benzene-1,4-dithiolate molecule \cite{reed97}, 
buckminsterfullerene \cite{joachim97}, individual atoms \cite{scheer98},
nanoclusters \cite{wildoer98} and carbon nanotubes \cite{martel98}.
These initial observations of conduction in molecules were followed 
quickly by experimental demonstration of the simplest devices: 
molecular diodes \cite{metzger2000}, switches \cite{chen99} 
and memory elements \cite{reed2001} although  molecular equivalent of 
a transistor is yet to be discovered.

Several theoretical approaches have been recently developed to describe
electronic conductivity of  molecules \cite{nitzan2001}. 
One of the widely used methods is based on the
combination of extended H\"uckel model or density functional 
theory to treat the molecular electronic structure, an  
electronic reservoir model for the contacts
and linear response approximation to compute the electronic current as 
a function of the applied voltage 
\cite{mujica94,tian98,hall2000,gutierrez2001,derosa2001}. 
The most recent applications used Lippmann-Schwinger or non-equilibrium 
Green's functions
formalisms plus ``jellium'' model for the contacts and 
for the molecular-contact interactions \cite{diventra2000,taylor2001}. 
In all these approaches, the system is partitioned into three parts: 
molecular wire and two electron reservoirs to represent metallic contacts.
The use of the model dependent reservoir-molecule interaction as one of the 
most pivotal ingredient diminishes the ability of these  methods and 
prevents from a development of predictive and quantitative techniques for 
the electronic transport problem. Recent density functional theory based
calculation demonstrates that a molecular
wire is bonded to a gold contact by the thiol bridge which is 
known to be the strong covalent interaction \cite{johansson2000}. 
The empirically treated 
coupling between metal contacts and the molecular wire, 
which serves as the interaction for the linear response calculations and 
represents the  kernel of the corresponding Lippmann-Schwinger equation,  
is at least strong as intermolecular interaction. 
Recent papers also indicate the exact nature of the molecule-contact 
bonding is critical in predicting the correct order of magnitude 
for the current with applied bias  \cite{diventra2000}. 

The miniaturization shrinks the size of the system to a level
when the problem can be addressed by ab initio electronic structure 
methods. As an example consider a standard prototype molecular device 
system with  a single benzene-1,4-dithiolate molecule
covalently bonded to two  gold electrodes \cite{reed97}. 
The problem can be reduced 
to the system where the molecule is bonded to two  
small Au clusters \cite{derosa2001}. 
A gold cluster with reasonably small edge influence on the 
molecular-surface bonding contains 10 atoms \cite{johansson2000}. 
With a  single benzene-1,4-dithiolate molecule as the molecular 
wire the contact-wire-contact system consists of just 32 atoms. 
Systems of such size can be routinely simulated  by standard 
quantum chemistry codes. The main obstacle for the application of the 
first principle electronic structure methods is not the size of the system 
but modeling of the electronic current and interpretation of the
applied voltage.
Suppose that we have solved Hartree-Fock or Kohn-Sham equations 
for a  molecular device. The ground state wave function is real 
in absence of a magnetic field
and therefore  the current density computed from this wave function is zero
everywhere in the sample.
Another important issue is an interpretation of the applied 
voltage.  Virtually all present theoretical approaches are 
pivotally relied on simplified, noninteracting electrons  type models 
for the metal contacts where the external voltage bias can be 
straightforwardly associated with the Fermi energies shift between 
left and right electronic baths. One of the reasons for 
use  electronic reservoir  models to represent the metallic 
contacts is principle theoretical difficulties related to a 
rigorous quantum mechanical 
definition of the applied voltage. 

In this paper, we advocate an alternative approach 
where the use of the electronic reservoir to model 
contacts is completely avoided from the very beginning and a quantum system 
with current is treated effectively as a {\it closed }
system. We begin by describing the variational principle for 
a quantum system with current. We then derive 
the Schr\"odinger equation with probability flux
which provides exact wavefunctions on the manifold 
of a given current density. 
We next specify two-terminal molecular device  steady 
current subsidiary condition and develop the Schr\"odinger 
equation for a two-terminal molecular device. 
We discuss the use of the energy ballance condition to
compute current-voltage characteristics of quantum systems. 
We then derive the Landauer formula for non-interacting
charge carriers within the formalism.
Finally, we demonstrate the salient features of the approach 
through the numerical solution of the self-consistent 
Schrodinger equation for one-dimensional harmonic 
oscillator with constrained current.

\section{Schr\"odinger equation with probability current}

\subsection{Variational principle}

We start with the nonrelativistic Hamiltonian for a particle in the
external scalar potential $v({\bf r})$ 
(atomic units are used throughout the paper
unless otherwise mentioned): 
\begin{equation}
H=-\frac{1}{2}  {\nabla}^{2} +v({\bf r}) \; .
\label{hamiltonian}
\end{equation}
We define density and paramagnetic current density as:
\begin{eqnarray}
\rho({\bf r}) &=& 
\psi   ({\bf r})  
\psi^* ({\bf r} ) \\
\nonumber
 \\
{\bf j}({\bf r})&=& \mbox{Im}
\left\{  \psi^* ({\bf r} ) 
\nabla \psi({\bf r} )
\right\} 
 \label{j}
\end{eqnarray}
where $\psi({\bf r})$ is not generally an  eigenvector of the Hamiltonian 
$H$. We will specify an eigenproblem for the $\psi({\bf r})$ 
in the next section. The physical current is equivalent to the paramagnetic 
current densities in absence of an external magnetic field.

An eigenvector for a bound state of a real Hamiltonian
can be always made real by a change of the phase of the wave function. 
The Hamiltonian $H$ is real therefore
if we just solve standard Schr\"odinger equation with the Hamiltonian $H$,
we will obtain real wave function $\psi({\bf r})$ and 
the current density ${\bf j}({\bf r})$ will be automatically zero. The
standard way to get a complex wavefunction out of a real Hamiltonian is 
to assume certain boundary conditions. These boundary conditions could be
an assumption of an incoming plane wave in scattering theory 
or periodic boundary conditions for a translationally invariant system. 
To solve a scattering problem for a molecular device is possible only at the
cost of major unphysical approximations, e.g. via considering the 
molecular device as a singe scattering center. 
Another approach is to assume periodic boundary conditions.
The periodic boundary conditions prevent one from studying 
any situations where the change of chemical potential across the system is  
of finite magnitude because the chemical potential must be also periodic 
in space. The local chemical potential is the decreasing function in 
the direction of the current
and therefore the system can not be treated as translationally 
invariant at least along 
the current axis.

We  have developed an  alternative approach where a current carrying quantum system 
is treated effectively as a {\it closed } system with a subsidiary condition for the 
generally discontinuous current density distribution.  We start our derivation with 
the assumption that  current carrying states of an open quantum system
can be described in terms of wave functions of an effectively closed quantum system
if the current is maintained via explicit constraint. 
This assumption is a simplified, particular case of the standard assumption in 
statistical mechanics:  a system coupled to a thermal bath can be described 
with statistical ensembles of a closed system in which the system-bath interaction is 
not explicitely considered, but enters only via Lagrange multipliers and subsidiary 
constraints. It has been recently demonstrated that a steady current open mesoscopic  
system  can be described in terms of an ensemble of carrier states 
in the closed mesoscopic device itself if the steady current is additionally 
constrained \cite{heinonen93,johnson95}.

We require that the current density is constrained to be specified function
${\bf I}({\bf r}) $. It  results in the subsidiary constraint equation:
\begin{equation}
{\bf j}({\bf r})- {\bf I} ({\bf r})=0  \; .
\label{constraint}
\end{equation}
The expression (\ref{constraint}) generally can not be obtained by 
differentiating of a functional of $\psi$ and $ \psi^* $
with respect to $ {\bf r} $. It means that the constraint eq.(\ref{constraint})
is nonholonomic \cite{courand}. The variational problem  is to minimize 
the total energy with respect to  $\psi$ subject to the nonholonomic constraint 
for the current density (\ref{constraint}). Two types of treatment 
are possible: we can either choose $\psi({\bf r})$ in which the constraint is 
implicit or we can impose the constraint explicitely 
by using the Lagrange multiplier. 
We will follow the second approach which was 
recently suggested by Kosov and Greer \cite{kosov2001}, 
elimination the part of the wavefunction by imposing 
the explicit constraint. Although the  constraint formulated as eq.(\ref{constraint}) 
is vector and nonholonomic it can be  included 
into a variational functional via pointwise, vector Lagrange multipliers 
${\bf a}({\bf r})$ \cite{kosov2001}.

Within the Lagrange multiplier approach the constraint is explicitly 
achieved via introduction of the auxiliary functional:
\begin{equation}
\Omega[ \psi]= 
\langle \psi |H|\psi \rangle - 
E \left( \langle \psi |\psi \rangle  -1\right) - 
\int {\bf a}( {\bf r}) \cdot \left({\bf j}({\bf r})- 
{\bf I}({\bf r})\right) d{\bf r }  \; .
\label{omega}
\end{equation}
The first two terms are standard in the variational derivation of the 
Schr\"odinger equation 
with the first giving the total energy and the second is introduced to 
maintain the normalization of the wave function.
The third term with the vector Lagrange multipliers 
${\bf a}({\bf r})$
has been introduced to impose the subsidiary constraint for
the current density.

A use of the variational principle can be justified only for closed quantum
systems.
The choice of ${\bf I}({\bf r}) $ can not be obviously completely arbitrary
to maintain the closed system boundary condition. 
The localized in space quantum system will be considered closed if, first,
there are no in- and out- probability flows in  the system;
and, second, if the wave function $\psi$
is an eigenvector of a Hermitian operator (it prevents the complex 
eigenenergies and therefore effectively closes the system).
We will demonstrate in the next section that both these conditions are satisfied 
within our approach.

\subsection{Self-consistent Schr\"odinger equation}

The minimization of the total energy subject 
to the subsidiary constraint for the current density (\ref{constraint}), i.e
the variation of the auxiliary functional $\Omega[\psi] $,  results into the following 
non-linear Schr\"odinger equation:
\begin{equation}
\left(
-\frac{1}{2} \Delta
+ v({\bf r})   - \frac{1}{2i} \left[ {\bf a}({\bf r}), \nabla \right]_+ \right) \psi({\bf r})= 
E \psi ({\bf r}) \; .
\label{schr1}
\end{equation}
The anti-commutator term,
\begin{equation}
\frac{1}{2i}\left[ \nabla, {\bf a}({\bf r}) \right ]_+ = 
\frac{1}{2i}\left(\nabla  {\bf a}({\bf r}) 
+ 2{\bf a}({\bf r})  \nabla \right),
\end{equation}
is the additional completely imaginary potential arising 
directly from the constraint on the 
current density 
(consult with appendix A for the corresponding functional derivatives ). 
This additional constraint potential forces the wave function 
$\psi({\bf r}) $ to be irremovably complex and insures that  
$\psi({\bf r})$ produces the required current density ${\bf I}({\bf r}) $. 

A physical interpretation of the Lagrange multiplier $ {\bf a }({\bf r})$ 
can be made more explicit if 
we re-write eq.(\ref{schr1}) in the following form:
\begin{equation}
\left[
\frac{1}{2} \left(- i \nabla -  {\bf a}({\bf r})
\right)^2 
+ \frac{1}{2} {\bf a}({\bf r}) \cdot  {\bf a}({\bf r})+
  v({\bf r})  \right] \psi({\bf r})= 
E \psi ({\bf r}) \; .
\label{schr2}
\end{equation}
It can be now established that the current constraint has introduced the terms 
which are mathematically equivalent to an external vector potential. But unlike an 
external vector potential,  
the Lagrange multiplier ${\bf a}({\bf r})$
does depend upon the  wave function $\psi({\bf r})$ 
and eq.(\ref{schr2}) becomes nonlinear and must be solved self-consistently.
The eigenvalue problem eq.(\ref{schr2}) is generated by Hermitian operator since 
the Lagrange multiplier ${\bf a}({\bf r})$ is real. 
Being a solution of the Hermitian eigenproblem
the eigenenergy $E$ is real and the wavefunctions $\psi$ form a complete set in Hilbert space.
Within our approach we deal with
the localized in space system with the wave functions generated by the hermitian 
eigenproblem (\ref{schr2}). Such system is by definition closed and therefore the use 
of the variational principle is formally justified.

The  Schr\"odinger equation with current (\ref{schr2}) is not yet in a form 
allowing for a solution to be found as the Lagrange multiplier  
${\bf a}({\bf r})$ is not known yet.
We next seek a solution for the Lagrangian multiplier
${ \bf a}({\bf r})$. 
In the developing of this method we have found out that the most significant source 
of numerical instabilities in the solution of the nonlinear Schr\"odinger equation 
(\ref{schr2}) arises from the determination of the Lagrange multiplier ${\bf a}({\bf r})$.
The described below procedure obtaining 
of the Lagrange multiplier has been found to be the most robust.
Multiplying eq. (\ref{schr1}) from the left on the  $\psi({\bf r})^*$
and subtracting from the obtained equation its own Hermitian conjugate, 
yields
\begin{equation}
\nabla \cdot {\bf j}({\bf r}) = \nabla \cdot ( { \bf a}({\bf r}) \rho({\bf r})) \;.
\label{lagrange1}
\end{equation}
Equation (\ref{lagrange1}) can be resolved for ${\bf a}({\bf r})$ with the 
assumption that both the density  $\rho({\bf r})$ and the current density 
${\bf j}({\bf r})$ decay to zero at infinity. In this case the direct 
integration of eq.(\ref{lagrange1}) results into the following expression 
for the Lagrange multiplier ${\bf a}({\bf r})$:
\begin{equation}
{ \bf a}({\bf r}) = \frac{{\bf j}({\bf r})}{\rho({\bf r})}
\label{lagrange2}
\end{equation}
Substitution of the constraint for the current density eq.(\ref{constraint})
into eq.(\ref{lagrange2}) yields the simple expression in which the 
Lagrange multiplier depends upon the wave function only via the density 
in the denominator:
\begin{equation}
{ \bf a}({\bf r}) = \frac{{\bf I}({\bf r})}{\rho({\bf r})}
\label{lagrange3}
\end{equation}

There is a direct similarity between equation for the Lagrange multiplier
(\ref{lagrange2}) and the definition of the velocity of a quantum trajectory within
de Broglie-Bohm causal description of quantum mechanics \cite{holland}. Within the 
de Broglie-Bohm dynamics the velocity ${ \bf u}({\bf r},t)$ of a trajectory at a given point 
is computed as \cite{bittner2000}:
\begin{equation}
{ \bf u}({\bf r},t) = \frac{{\bf j}({\bf r},t)}{\rho({\bf r},t)} \; ,
\label{velocity}
\end{equation}
therefore, at least for stationary systems the vector Lagrange 
multiplier ${ \bf a}({\bf r})$  can be directly related 
to the velocity for a corresponding de Broglie-Bohm trajectory.  
By this analogy we can make the physical interpretation of the
Lagrange multiplier more explicit. In the de Broglie-Bohm quantum
dynamics to maintain the constant flux at the given density profile 
one needs to adjust the local trajectory velocities. 
The vector Lagrange multiplier appears to play 
the simular role in the Schrodinger equation with constrained current.
Likewise for a point with the non-zero trajectory velocity, 
the current density has finite value at a 
point with the non-zero  vector Lagrange multiplier.
And likewise the  de Broglie-Bohm velocity,
a product of the vector Lagrange multiplier on the probability density 
yields the current density.

With this choice of the Lagrange multiplier (\ref{lagrange3})
we obtain the Schr\"odinger equation with the current ${\bf I}({\bf r})$
in the self-consistent and closed form:
\begin{equation}
\left[
\frac{1}{2} \left(- i \nabla -  \frac{{\bf I}({\bf r})}{\rho({\bf r})} \right)^2 
+ \frac{1}{2} \frac{ I^2 ({\bf r})}{\rho^2({\bf r})}+
  v({\bf r})  \right] \psi({\bf r})= 
E \psi ({\bf r}) \; .
\label{schr3}
\end{equation}
The derivation of the self-consistent  
Schr\"odinger equation eq.(\ref{schr3}) composes one of the central results of the paper.

\section{Schr\"odinger equation for a two terminal molecular device}

The desire to develop the basic Schr\"odinger
equation for calculations of  $I-V$ characteristics of two terminal molecular devices
is the impetus for the present study. To achieve
this aim, we specify the molecular wire constraint on the current density:
\begin{equation}
\int dy\, dz\, j_x ({\bf r}) = I_x \; \theta(x-L,R-x)  \;,
\label{constraint_wire}
\end{equation}
where $I_x$  is the steady current through the molecular device. The step function,
\begin{eqnarray}
\theta(x-L,R-x)= \theta(x-L) \theta(R-x)  \; ,
\end{eqnarray}
 equals to 1 if $L\le x \le R$ and zero otherwise.
$L$ and $R$ are the left and right boundaries for the molecular device. 
Within this description, net current flow is aligned along
the $x-$axis and there are no in- and out flow: the current starts at the left boundary L
and completely absorbed at the right R.
With this constraint it is not required to fix the 
full vector ${\bf j}({\bf r})$ everywhere in the sample.
Only the net current flow across a cross section
$\int dy\, dz\, j_x({\bf r})$ is constrained, and this
quantity is readily available experimentally.
This is a simple geometric arrangement to specify current flow for the two terminal molecular
device and can be extended to more complex device topologies, e.g. for the 
three terminal devices where three tips are connected to a single molecule
\cite{diventra2000_b}. 

Likewise, as for the derivation of the exact  Schr\"odinger 
equation in the preceding section, 
with this choice of the constraint (\ref{constraint_wire}) the minimization 
of the auxiliary functional (\ref{omega}) now yields the 
Schr\"odinger equation for the steady current two terminal molecular device:
\begin{equation}
\left(
-\frac{1}{2} \Delta
+ v({\bf r})   - 
\frac{1}{2i} \left[ a_x( x), \frac{\partial}{\partial x} \right]_+ \right) \psi({\bf r})= 
E \psi ({\bf r}) \; .
\label{schrwire1}
\end{equation}

The expression for the Lagrange multiplier given by eq.(\ref{lagrange3})
can be straightforwardly extended to the steady current 
molecular device,  with the result is 
\begin{equation}
a_x(x) = \frac{I_x}{\rho_{yz}(x)} \theta(x-L,R-x) \;,
\end{equation}
where we have introduced the quantity
\begin{equation}
\rho_{yz}(x) = \int dy dz \rho ({\bf r}).
\end{equation}

With the expression for the Lagrange multiplier  $a_x( x)$ in hands, we can complete the
Schr\"odinger equation for the two-terminal molecular device with steady current $I_x(x)$
\begin{equation}
\left(
-\frac{1}{2} \Delta
+ v({\bf r})   - \frac{I_x}{2i} \left[\frac{\theta(x-L,R-x) }{\rho_{yz}(x)}, 
\frac{\partial}{\partial x} \right]_+ \right) \psi({\bf r})= 
E \psi ({\bf r}) \; .
\label{schrwire2}
\end{equation}
Some aspects of eq.(\ref{schrwire2}) deserve special discussion.
The imaginary potential which enforces the wave function to 
produce steady current $ I_x $
depends upon the $x$ coordinate only. The range of this potential is 
restricted by the molecular device boundaries. One of the terms in this 
potential is proportional to the derivative of the step function,
\begin{equation}
\frac{\partial}{\partial x} \theta(x-L,R-x) =\delta(x-L) -\delta(x-R)\; ,
\end{equation}
and therefore yields the singular imaginary $\delta$-function potential on
the device boundaries. It is known from the standard quantum mechanics 
textbooks \cite{fluege71} that a $\delta$-function potential results 
in the discontinuity of the first derivative of the corresponding solutions
of the Schr\"odinger equation. 
Likewise, the imaginary component of the
eigenfunction $\psi({\bf r})$ of the Schr\"odinger equation for a
molecular device (\ref{schrwire2})    has the  discontinuous first derivative 
at the points $x=L$ and $x=R$.

The final comment regarding the computation of the I-V characteristic of a 
two-terminal molecular device is in due order. The current carrying states
obtained as the solutions of the molecular device Schr\"odinger equation  
(\ref{schrwire2}) must now be associated with current-voltage 
characteristics available 
experimentally. The rigorous definition of the applied voltage
is a controversial issue if one does not invoke to noninteracting electron 
reservoirs to represent contacts. We propose the following macroscopic definition of the 
applied voltage which has the direct relation with the quantities 
measured experimentally. 
The energy increase from the applied voltage $U$ is 
computed by the numerical integration with an assumption of  linear 
voltage drop between the left and right boundaries:
\begin{equation}
\Delta E_d(U) =\frac{U}{R-L} \int \limits_L^R dx  (R-x)\rho_{yz}(x) \; .
\end{equation}
We compute the energy of the non-active, i.e. with zero current, molecular 
device and compute the total energy of the same molecular device at an  
experimentally given current $I_x$.
Then the energy increase for the establishment of 
the current carrying state, i.e. the energy difference 
\begin{equation}
\Delta E_d(I_x) =E_{d}(I_x)- E_{d}(I_x=0)
\end{equation}
is calculated. Because  of the energy ballance condition  
the $\Delta E_d(U)$ must be equal
to the $\Delta E_d(I_x)$ and it results into the following equation for the applied 
voltage bias:
\begin{equation}
U= 
\frac{(R-L)  }{\int \limits_L^R dx (R-x) \rho_{yz}(x)}  \Delta E_d(I_x)\; .
\end{equation}
 The eigenenergy $E$ of the Schr\"odinger equation (\ref{schrwire2})
should not be confused with the energy of molecular device. The eigenenergy $E$ 
serves only as the Lagrange multiplier to enforce orthogonality of the wave function.
The energy of the molecular device can be obtained as an expectation value of
the real physical Hamiltonian $H$ with the integration
restricted by the molecular device region:
\begin{equation}
E_d(I_x) =\int dy dz \int \limits_{L}^{R} dx 
\psi^{*} ({\bf r}) \left( -\frac{1}{2} \Delta +v({\bf r}) \right) \psi({\bf r})\;,
\label{edevice}
\end{equation}
where $\psi({\bf r})$ is the solution of the Schr\"odinger equation (\ref{schrwire2})
with the given  $I_x$. Since the wavefunction $\psi({\bf r})$ is complex 
and the integration (\ref{edevice}) is not performed over whole space, it
is not obvious from the expression (\ref{edevice}) that the device energy $E_d(I_x)$ 
must be real. We will show now that the device energy $E_d(I_x)$ is real
if the wavefunction $\psi ({\bf r})$ produces the constrained 
current density (\ref{constraint_wire}). 
To prove this we subtract from eq.(\ref{edevice}) its own complex conjugate
and after a bit of algebra arrive to the following expression:
\begin{equation}
E_d(I_x)- E_d(I_x)^* = 
-i \int dy dz \int \limits_L^R dx \nabla \cdot {\bf j} ({\bf r}) = 
 -i \oint_S dS {\bf n} \cdot {\bf j} ({\bf r})\; .
\end{equation}
 The length of the quantum wire is aligned along
the $x-$axis and the surface integral does the integration over the 
surface of the rectangular box with one $y-z$ face at $x=L$, another 
 $y-z$ face at $x=R$ and any of the $x-y$ or $x-z$ faces are removed
at the infinity.
Within our constraint, there is a positive current
flowing into the $y-z$ face at $x=L$ and exactly the same positive
current flowing out of the $y-z$ face at $x=R$. There is no current flow 
out of the region in any of the $x-y$ or $x-z$ faces. Therefore
the surface integral vanishes 
\begin{equation}
\oint_S dS {\bf n} \cdot {\bf j} ({\bf r}) =0
\end{equation}
and the device energy $E_d(I_x)$  defined via eq.(\ref{edevice}) becomes real.

\section{Example calculations}

\subsection{Conductance of an ideal one-dimensional lead}

In order to establish connection with standard approaches 
and test the method we derive the Landauer formula \cite{landauer70}
using the Schr\"odinger equation for current carrying states
 eq.(\ref{schr3}) as the starting point.
Consider noninteracting electrons moving along a one-dimensional wire without any
scatterer. The electronic density does not depend upon $x$ yielding the following
Schr\"odinger equation with steady current $I$:
\begin{equation}
\left[
- \frac{1}{2 }\frac{d^2}{dx^2}+ 
i\frac{ I}{\rho}\frac{d}{dx}  \right] \psi_k(x)= 
E_k \psi_k (x) \; .
\label{schr4}
\end{equation} 

Equation (\ref{schr4}) has the plane-wave solution 
\begin{equation}
\psi_k(x) \sim \exp(i k x)
\end{equation}
with eigenenergy
\begin{equation}
E_k =\frac{k^2}{2 } - \frac{I k}{ \rho} \; .
\label{ek}
\end{equation}
The similar dispersion relation as given by eq.(\ref{ek}) has been 
obtained within the same approach for uniform electron gas with 
applied direct current \cite{kosov2001}.
For the small voltage and temperature transmissions of electrons 
take place only through the states in close vicinity of 
the Fermi level. There are  two vectors k, one is positive and one is negative,
which correspond to the Fermi energy:
\begin{eqnarray}
k_+ &=& \frac{I}{ \rho} +
\sqrt{ 2 E_F +\left(\frac{I}{ \rho}\right)^2 } \; ,
\label{k+} 
\\
k_- &=& \frac{I}{\rho} - 
\sqrt{ 2 E_F +\left(\frac{I}{\rho}\right)^2 } \; .
\label{k-} 
\end{eqnarray}
Following the Landauer approach we associate the voltage drop (at the zero temperature) 
with the gap between the singe-electron energies of the electrons
moving in the direction of the current ($k>0$) and electrons moving 
in the opposite direction ($k<0$):
\begin{equation}
 U = \frac{k_{+}^2}{2 } - \frac{k_{-}^2}{2 } \; .
\label{voltage1}
\end{equation}
Substituting the expressions for the $k_+$ and $ k_-$ eqs.(\ref{k+},\ref{k-}) into the 
Landauer's definition of the applied voltage (\ref{voltage1}), we readily find
\begin{equation}
I= G  U \;, 
\label{landauer1}
\end{equation}
with the conductance
\begin{equation}
G= \frac{\rho}{2 \sqrt{ 2 E_F + \left(\frac{I}{\rho}\right)^2}} \; .
\label{conductivity1}
\end{equation}
Then, the use of the standard relation between the Fermi momentum $k_F$ and the 
electronic density $\rho = k_F / \pi $ 
along with the small current  expansion of
the square root eq.(\ref{conductivity1}) results into the Landauer
formula  for a single transport channel of an ideal one-dimensional lead \cite{datta95}:
\begin{equation}
G= \frac{1}{2 \pi  } \;.
\label{landauer2}
\end{equation}
If we take into account the spin degeneracy of real electrons, the conductance
 becomes multiplied by 2. By accounting the degeneracy and converting the conductance
eq.(\ref{landauer2}) to SI units the standard value 
$G= e^2/ (\pi \hbar) = (12.9 \; K \Omega)^{-1}$ is obtained. Re-deriving the standard result
we demonstrate that in the low current and 
zero temperature limits our method is equivalent to the traditional approaches which
are based on the Landauer formula.

\subsection{Numerical solution of the Schr\"odinger equation with current}

We demonstrate applications of the method with numerical examples.
The numerical solution of a self-consistent Schr\"odinger equation is often performed 
within a finite basis set expansion for the wave function with subsequent diagonalization 
of the Hamiltonian matrix until the self consistency is reached. 
We have found out that
it is computational convenient to form the complete, real basis set from the
solutions of the zero-current Schr\"odinger equation:
\begin{equation}
\left(
-\frac{1}{2} \Delta + v({\bf r}) \right) \phi_{\mu}({\bf r})= 
\varepsilon_{\mu} \phi_{\mu} ({\bf r}) \; .
\label{schr_0}
\end{equation}
Then we expand the wave function of a current currying state on this basis
\begin{equation}
\psi({\bf r}) = \sum_{\mu} C_{\mu} \phi_{\mu} ({\bf r}) \;,
\label{expansion}
\end{equation}
the coefficients $C_{\mu} $ must be complex for the wave function 
which produces a  non-zero current.
With the expansion (\ref{expansion}) we arrive to the following complex, Hermitian
eigenproblem:
\begin{equation}
\sum_{\mu} \left( \varepsilon_{\mu} \delta_{\mu \nu} - i w_{\nu \mu}({\bf I}) \right) C_{\mu}
= E C_{\nu} \; ,
\label{eigen}
\end{equation}
where the current dependent matrix element $w_{\nu \mu}({\bf I})$ is
computed in appendix B and is given by the following integral:
\begin{equation}
w_{\nu \mu}({\bf I}) =\frac{1}{2} \;
\int d{\bf r} \;  \frac{{\bf I({\bf r})}}{\rho({\bf r})} \cdot
\left( \phi_m ({\bf r}) \nabla \phi^*_n ({\bf r}) 
- \phi^*_n ({\bf r}) \nabla \phi_m ({\bf r}) \right)\; .
\label{w_j}
\end{equation}
The matrix elements $w_{\nu \mu}({\bf I})$ are real and depend on the expansion coefficients 
$C_{\mu}$ via the density $\rho({\bf r})$ thereby making the eigenvalue problem 
(\ref{eigen}) nonlinear. 
The eigenproblem (\ref{eigen}) is  Hermitian because the matrix element 
$w_{\nu \mu}({\bf I}) $
changes its sign when we swap $\mu$ and $\nu $ indeces:
\begin{equation}
w_{\nu \mu}({\bf I}) = - w_{\mu \nu}({\bf I}) \; .
\end{equation}

The numerical calculations were carried out for the one-dimensional system.
With an eye on molecular device simulations
we solve the self-consistent Schr\"odinger equation for
the step function constraint  as a primary example:
\begin{equation}
j(x)=\; I \;\theta(x-L,R-x) 
\label{constraint_1d}
\end{equation}
The external potential was taken in the harmonic oscillator form and
we put $L=-2.0 $ and $R=2.0 $ as left and right boundaries for the
current carrying part of the system in all our calculations. 
Our basis set was built up from the $25$ eigenvectors of the zero current
Hamiltonian. The self-consistency was assumed to be achieved when the root 
mean square deviation of the wave function vectors of the two 
subsequent iterations was less then $10^{-8}$.

First, we plot the Lagrange multiplier as a function of the coordinate
on the figure 1.
The Lagrange multiplier is discontinuous  at 
the system boundaries. The Lagrange multiplier is vanishing out of the system 
region and it is the inverse function of the density $\rho(x)$ inside 
the system. 
As it can be expected from the analogy with the  de Broglie-Bohm quantum 
hydrodynamics (\ref{velocity}) (the lower the density, the larger the
local velocity)
the Lagrange multiplier reaches its maximum values in the regions 
of depleted density and has the minimal value at the point
of maximal  density.

In our next example we studied the deviation of the current carrying
wave function   $\psi ({\bf r})$ ($I=0.1$)
 from the zero-current ground state wave function $\phi_0 ({\bf r})$. 
The real component is depicted on the upper panel of the fig.2.
The  shape and magnitude of the real component do not undergo significant transformation 
with the establishment of the current carrying state. 
Although the admixture of the higher terms 
in the expansion (\ref{expansion}) results in the small damping oscillations 
on the tails of the real component of the wave function.
The imaginary part of the wave function ( plotted on the lower panel of the fig.2)
is identically zero when the current is vanished. 
The imaginary component reaches its negative minimum 
at the left boundary $L=-2.0$
of the current carrying part of the system and then it discontinuously changes 
the sign of the first derivative and undergoes almost linear increase
until it reaches its maximum at the right boundary $R=2.0$. When the 
coordinate $x$ goes out of the non-zero current region the
imaginary part of the wave function rapidly decays to zero value.

The establishment of the imaginary part of the wave function 
directly indicates the development of the current carrying state. 
We studied transformations of the imaginary component as a function  of 
the increased current. The imaginary components of the wave functions are plotted 
on the fig. 3 for three representative  steady currents: 
$I=0.01$, $I=0.5$ and $I=0.1$. 
At the very small current $I=0.01$ the imaginary wave 
function is almost zero everywhere except the only small negative minimum at
the left boundary and with the positive maximum at the right. 
The magnitudes of the imaginary component at the extremums are
nearly linearly proportional to the value of the constrained current.
The slope of the imaginary component, i.e. the derivative, is always positive 
and almost constant between the left and right boundaries of the system.
The imaginary part becomes zero when the real component reaches its 
maximum, i.e. at the point $x=0.0$. Although the total wave function is 
real at the point  $x=0.0$ the current is not vanished because the
derivative of the wave function remains complex.

Now we turn to the calculations of a ``device'' energy required to establish
the current carrying state. 
Being the Lagrangian multiplier to maintain the wavefunction orthonormality, 
the eigenenergy $E$ of the eigenproblem (\ref{eigen}) is not directly related 
to the energy of the system. The energy increase due to the 
establishment of the current carrying state is defined via 
the space restricted expectation value of the Hamiltonian (\ref{edevice}):
\begin{equation}
E_d(I) =\int\limits^R_L dx \; 
\psi^{*}(x) \left ( - \frac{1}{2} \frac{d^2}{dx^2} +v(x) \right) \psi(x)
=\sum_{\mu \nu} C^{*}_{\mu} C_{\nu} \varepsilon_{\nu}  
\int\limits^R_L dx \;  \phi_{\mu}(x) \phi_{\nu}(x) \; .
\label{ed2}
\end{equation}
The table 1 contains the energy $E_d(I)$ and the energy difference between 
the nonactive, i.e. zero-current,  and current $I$ carrying state as a 
function of the increased current. 
The ``device'' energy is gradually increased with increasing current. 
As it can be deduced from the table 1, we need to pump into the system 
$\simeq 0.14 \; a.u.$ of energy to establish the current 
$I=0.1 \;  a.u. $ carrying state.
A simple fit of  the data from the table 1 provides only 
the quadratic and cubic  
dependences of the ``device'' energy upon the steady current (for $I\le 0.1$):
\begin{equation}
E_d(I)= E_d(I=0) + 17.4 I^2 -36.5 I^3 \; ,
\end{equation}
thereby  demonstrating that the 
term linear in the current has the minor importance for the energetics of 
the current carrying quantum system
and the main energy contribution comes from the  $I^2$ term.
This energy dependence  is generally agreed 
with the current-density functional theory of inhomogeneous 
interacting electron gas  where  the current dependent 
correction to the standard exchange-correlation 
energy functional is also mainly proportional 
to $I^2$  \cite{vignale87}. 

\section{Conclusions}
In this paper, we have given a variational formulation of the 
Schr\"odinger equation with non-zero current. The Schr\"odinger equation 
with current is derived via  a  constrained minimization of the
total energy 
with a subsidiary condition for the current density. The subsidiary condition 
for the current density is maintained during the course of variation
by a vector, pointwise Lagrange multiplier.
An explicit elimination of this Lagrange multiplier
results into the closed, self-consistent form of Schr\"odinger 
equation with current.
We showed that the current carrying states are eigenvectors of the 
complex, Hermitian operator. 
The formulation has been developed for general current density topologies
and then specified for the case of a two-terminal molecular device.
We showed how the energy ballance condition can be used for rigorous 
computation of  I-V characteristics of molecular devices: the energy increase due 
to the establishment of a current carrying state is associated with the 
applied voltage bias.  We
demonstrate the salient features of the approach 
by re-deriving the Landauer formula for the conductance of an
ideal one-dimensional lead
and through a 
solution of the one-dimensional Schr\"odinger equation
with fixed current density.
The complex, Hermitian matrix was formed on a   basis of the eigenvectors 
of the zero-current Hamiltonian.
The numerical solution was achieved via the self-consistent 
solution of the complex, Hermitian eigenproblem. 
We found out that a basis set formed from the zero current Hamiltonian 
eigenvectors provides a robust algorithm for the self-consistent convergence. 
Our method has the great compatability with standard electronic structure methods
which are also based on a variational principle
e.g. Hartree-Fock or configuration interaction approximations.

\acknowledgements 

The author wishes to thank  G. Stock 
for stimulating discussions, critical reading the manuscript 
and his continuous support of the project. The author also thanks
E.R.Bittner, J.C.Greer and A.Nitzan for valuable discussions.

\newpage
\appendix

\section{Functional derivatives of the constraint functional}

In this Appendix we compute the functional derivative of the
constraint functional.
The constraint functional has the following form
\begin{equation}
\Lambda[\psi] = 
\int d{\bf r} \; {\bf a} ({\bf r}) \cdot ({\bf j} ({\bf r}) -{\bf I}({\bf r})) \; .
\end{equation}
The direct variation of the constraint functional with respect to the
$\psi^* ({\bf r})$ yields:
\begin{eqnarray}
\frac{\delta \Lambda[\psi]}{ \delta \psi^* ({\bf r})} &=&
\int d{\bf r}'\; \frac{ \delta \Lambda }{\delta {\bf j}({\bf r}')} \cdot 
\frac{\delta {\bf j}({\bf r}')} { \delta \psi^* ({\bf r})}= 
\int d{\bf r}' \;{ \bf a}({\bf r}') \cdot \frac{\delta {\bf j}({\bf r}')} { \delta \psi^* ({\bf r})}
\nonumber
\\
&=&\frac{1}{2i} \int d{\bf r}'{ \bf a}({\bf r}') \cdot \frac{ \delta}{ \delta \psi^* ({\bf r})}
\left\{ \psi^*({\bf r}') \nabla \psi({\bf r}') - \psi({\bf r}') \nabla \psi^*({\bf r}') \right\}
\nonumber
\\
&=&\frac{1}{2i}  { \bf a}({\bf r}) \cdot \nabla \psi({\bf r}) -
\frac{1}{2i} \frac{ \delta}{ \delta \psi^* ({\bf r})} 
\int d{\bf r}' \; { \bf a}({\bf r}') \cdot \psi({\bf r}') \nabla \psi^*({\bf r}')
\nonumber 
\\
&=&\frac{1}{2i}  { \bf a}({\bf r}) \cdot \nabla \psi({\bf r}) -
\frac{1}{2i} \frac{ \delta}{ \delta \psi^* ({\bf r})} 
\int d{\bf r}' \; \nabla \cdot  \left\{{ \bf a}({\bf r}') \psi({\bf r}') \psi^*({\bf r}')\right\}
\nonumber 
\\
&+&\frac{1}{2i} \frac{ \delta}{ \delta \psi^*({\bf r})}
\int d{\bf r}' \;  \psi^* ({\bf r}') \nabla \cdot \left\{ { \bf a}({\bf r}') \psi({\bf r}')\right\}
\nonumber 
\\
&=&
\frac{1}{2i}  { \bf a}({\bf r}) \cdot \nabla \psi({\bf r}) -
\frac{1}{2i} \frac{ \delta}{ \delta \psi^* ({\bf r})} 
\oint dS \; {\bf n} \cdot { \bf a}({\bf r}') \psi({\bf r}') \psi^*({\bf r}')
+\frac{1}{2i} \nabla \cdot \{ { \bf a}({\bf r}) \psi({\bf r}) \}
\nonumber 
\\
&=&\frac{1}{2i}  { \bf a}({\bf r}) \cdot  \nabla \psi({\bf r})+
\frac{1}{2i} \nabla \cdot \{ { \bf a}({\bf r}) \psi({\bf r}) \} =
\frac{1}{2i} \left[ {\bf a}({\bf r}), \nabla \right]_+ \psi({\bf r})
\nonumber
\end{eqnarray}

\section{Matrix elements of the anti-commutator 
$ \left[ {\bf a}({\bf r}),\nabla \right]_+$}

The computation of the matrix element of the anti-commutator is presented below.
The notation $\nabla^{a}$ means that the gradient acts on the Lagrange 
multiplier  ${\bf a}({\bf r})$ only. 

\begin{eqnarray}
&& \langle \phi_n | \left[ {\bf a}({\bf r}),\nabla \right]_+ |\phi_m \rangle =
\langle \phi_n | \nabla^{a} \cdot {\bf a}({\bf r}) 
+ 2  {\bf a}({\bf r}) \cdot \nabla |\phi_m \rangle 
\nonumber
\\
&& =
\int d{\bf r} \; \phi^*_n ({\bf r})  \phi_m ({\bf r}) \nabla \cdot {\bf a}({\bf r}) + 
2 \int d{\bf r} \; {\bf a}({\bf r}) \cdot \phi^*_n ({\bf r})  \nabla \phi_m ({\bf r}) 
\nonumber
\\
&& =
\int d{\bf r} \; \nabla \cdot \{ {\bf a}({\bf r}) \phi^*_n ({\bf r})  \phi_m ({\bf r}) \}-
\int d{\bf r} \; {\bf a}({\bf r}) \cdot \nabla \{\phi^*_n ({\bf r})  \phi_m ({\bf r}) \}
+2\int d{\bf r} \; {\bf a}({\bf r}) \cdot \phi^*_n ({\bf r})  \nabla \phi_m ({\bf r}) 
\nonumber
\\
&& =
\oint dS \; {\bf n} \cdot  {\bf a}({\bf r}) \phi^*_n ({\bf r})  \phi_m ({\bf r}) -
\int d{\bf r} \; {\bf a}({\bf r}) \cdot \left( \phi_m ({\bf r}) \nabla \phi^*_n ({\bf r})  
- \phi^*_n ({\bf r}) \nabla \phi_m ({\bf r}) \right)
\nonumber
\\
&& =
 -
\int d{\bf r} \; {\bf a}({\bf r}) \cdot \left( \phi_m ({\bf r}) \nabla \phi^*_n ({\bf r})  
- \phi^*_n ({\bf r}) \nabla \phi_m ({\bf r}) \right)
\nonumber  
\end{eqnarray}

\clearpage


\newpage
\noindent
{\bf Table 1}. The energy of the active, i.e  current carrying $L \le x \le R$,
 part of the system as a function of the steady current.
   $\Delta E_d =E_d(I)-E_d(I=0)$ is the energy increase to establish of 
the  current  $I$ carrying state.

\vspace{3cm}
\noindent
\begin{tabular}{|c|c|c|}
\hline
~~~~$I\;\;\;\;$~~~~     & ~~~~$E_d(I)\;\;\;\;$~~~~   & ~~~~$\Delta E_d(I) \;\;$~~~~   
   \\
\hline
0.00        & 0.4977      & 0.0000
   \\
\hline
0.01        & 0.4994      & 0.0017
   \\
\hline
0.02        & 0.5043      & 0.0066
   \\
\hline
0.03        & 0.5124      & 0.0147
    \\
\hline
0.04        & 0.5233      & 0.0256
    \\
\hline
0.05        & 0.5368      & 0.0391
    \\
\hline
0.06        & 0.5526      & 0.0549
  \\
\hline
0.07        & 0.5706      & 0.0729
  \\
\hline
0.08        & 0.5905      & 0.0928
  \\
\hline
0.09        & 0.6122      & 0.1145
  \\
\hline
0.1         & 0.6356      & 0.1379
  \\
\hline
\end{tabular}

\newpage

\noindent
{\bf Figure 1}. The Lagrange multiplier $a(x)$ is plotted as a function of the coordinate
$x$. Calculations were performed for the  step function current density distribution
$ j=0.1\;  \theta(x-2,2-x) $.

\vspace{3cm}
\psfig{file=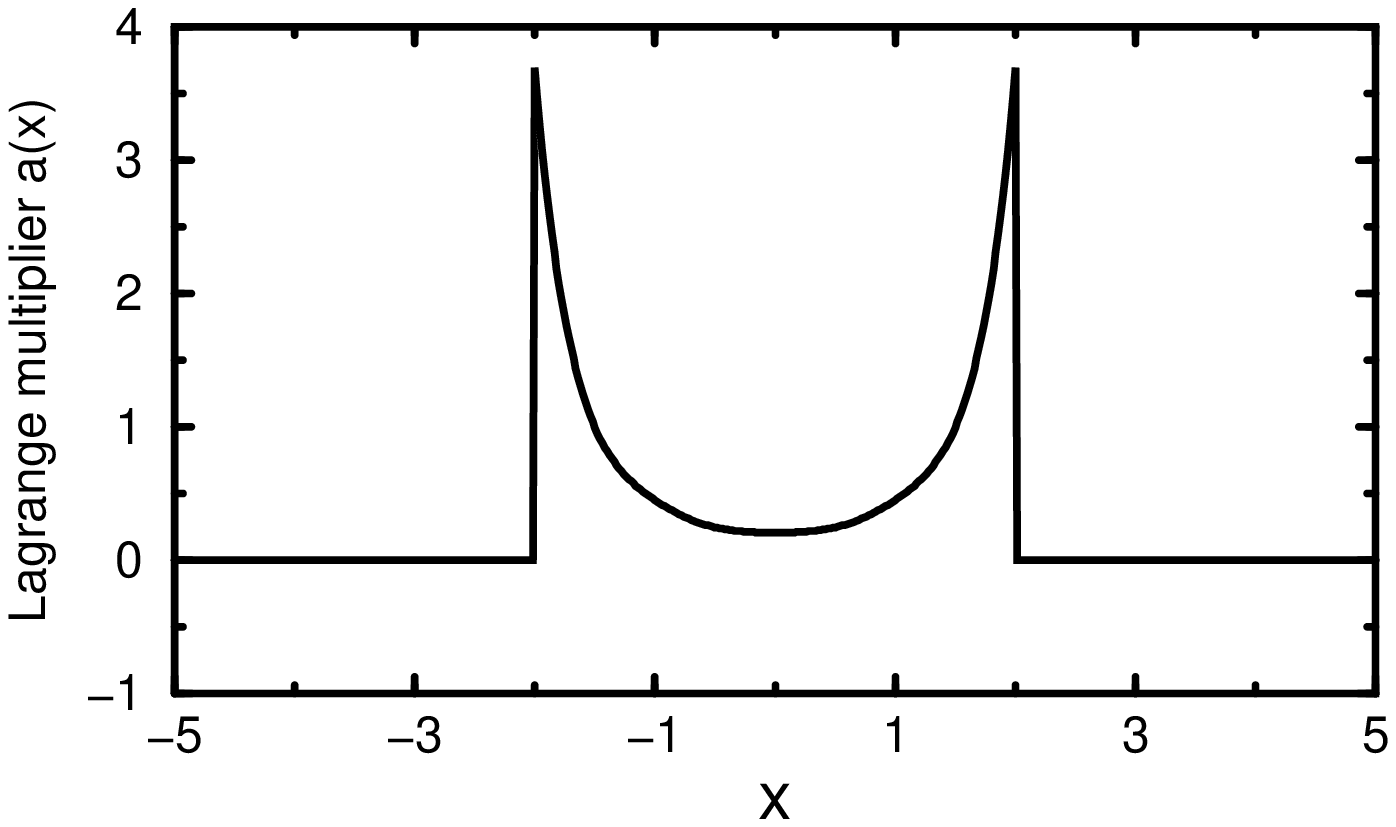,height=9cm}

\newpage

\noindent
{\bf Figure 2}. The spatial dependences of real and imaginary components of the
current carrying wave function.
Calculations were performed for the  step function current density distribution
$ j=0.1 \;  \theta(x-2,2-x) $. The dashed line on the upper panel is the ground state
wave function of the system with zero current.

\vspace{3cm}
\psfig{file=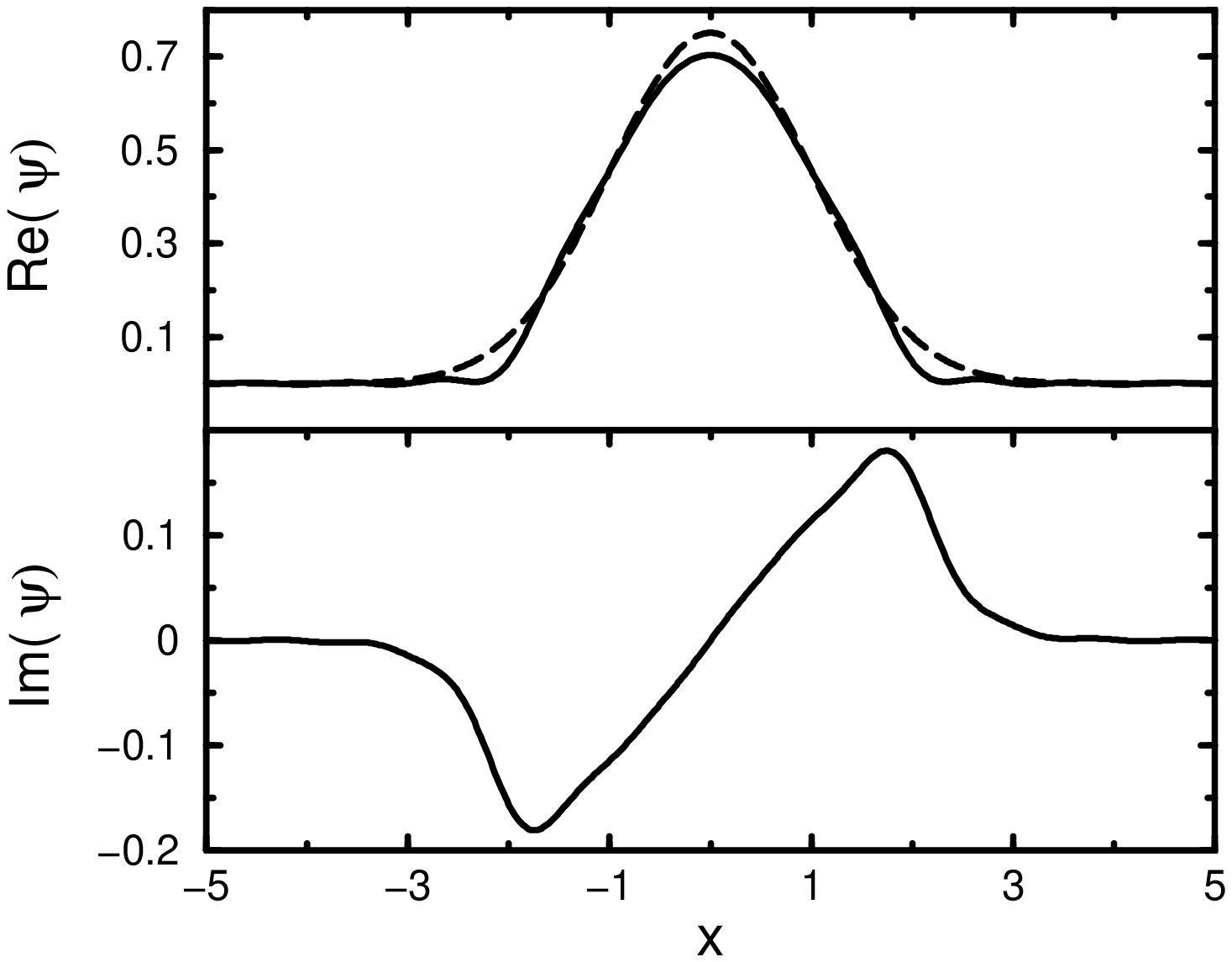,height=11cm}

\newpage

\noindent
{\bf Figure 3}. The spatial dependences of imaginary components of the wave functions
are plotted for different values of the current. The current density is constrained 
to be the step function $  j=I\;  \theta(x-2,2-x) $. The solid line corresponds 
to $I=0.01$, dotted to $I=0.05$ and dashed to $I=0.1$.

\vspace{3cm}
\psfig{file=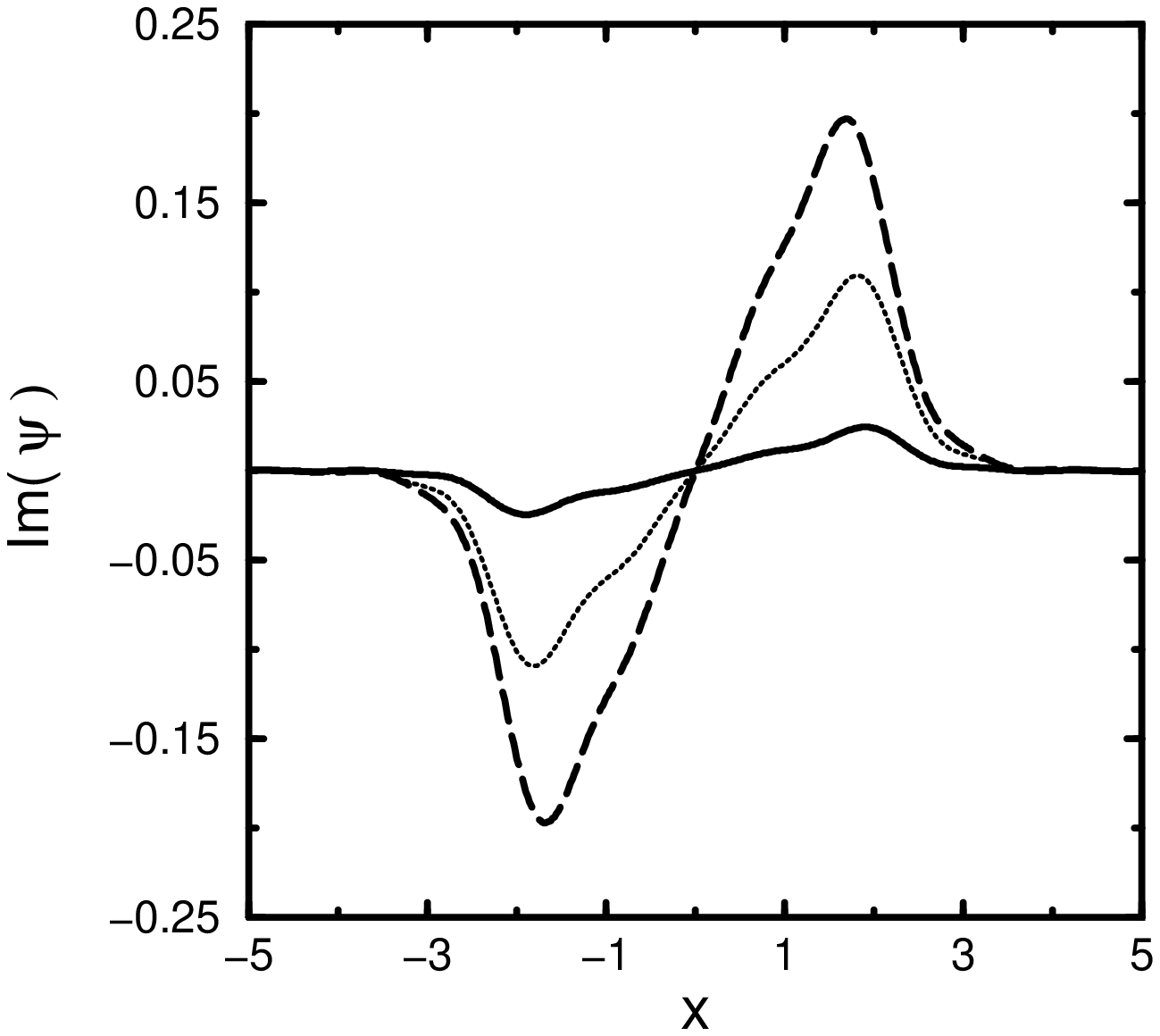,height=11cm}

\end{document}